\documentclass[10pt, journal]{IEEEtran}

\usepackage{lscape, subfigure, latexsym, amssymb, amsmath}
\usepackage[pdftex]{graphicx}
\usepackage{algorithm}
\usepackage{algorithmic}
\usepackage{graphicx}
\usepackage{epstopdf}
\usepackage{caption}
\usepackage[affil-it]{authblk}
\usepackage{subfigure}
\usepackage{amsmath}

\graphicspath{{figure/}}

\begin{document}

\title{Community Detection Algorithm Evaluation using Size and Hashtags}

\author{Paul Wagenseller III and Feng Wang
\thanks{Paul.Wagenseller@asu.edu, Feng.Wang.4@asu.edu}
}

\affil{ Arizona State University \\}

\maketitle

\thispagestyle{empty}

\begin{abstract}
Understanding community structure in social media is critical due to its broad applications such as friend recommendations, link predictions and collaborative filtering. However, there is no widely accepted definition of community in literature. Existing work use structure related metrics such as modularity and function related metrics such as ground truth to measure the performance of community detection algorithms, while ignoring an important metric, size of the community. \cite{Dunbar2016} suggests that the size of community with strong ties in social media should be limited to 150. As we discovered in this paper, the majority of the communities obtained by many popular community detection algorithms are either very small or very large. Too small communities don't have practical value and too large communities contain weak connections therefore not stable. In this paper, we compare various community detection algorithms considering the following metrics: size of the communities, coverage of the communities, extended modularity, triangle participation ratio, and user interest in the same community. We also propose a simple clique based algorithm for community detection as a baseline for the comparison. Experimental results show that both our proposed algorithm and the well-accepted disjoint algorithm InfoMap perform well in all the metrics. 
\end{abstract}

\section{Introduction}

%
%
%
%

Community is a fundamental element in social media for communication and collaboration. Understanding community structure in social media is critical due to its broad applications such as friend recommendations, link predictions and collaborative filtering. However, there is no widely accepted definition of community in literature. Informally, a community is a densely connected group of nodes that is sparsely connected to the rest of the network. In another word, a community should have more internal than external connections. Community detection problem has been extensively studied. Various algorithms have been proposed to minimize or maximize a goodness measurement, such as modularity, eigenvector, and conductance, etc. In general, the community detection algorithms can be categorized into disjoint and overlapping algorithms.  Comparison has been conducted to measure the performance of various community detection algorithms ~\cite{Santo2010}. Furthermore, ~\cite{Yang2012} validated the obtained communities with ground truth, which is the known community membership of users. 

However, previous research in community detection algorithm design and evaluation has ignored an important metric, size of the community. [1] suggests that the size of community with strong ties in social media should be limited to 150 due to the cognitive constraint and time constraint of human being in both traditional social network and Internet-based social network, i.e., social media. Therefore, too large communities contain weak connections therefore not stable. In addition, too small communities don't have practical value. In this paper, we use the term \emph{desirable community} to refer to the communities of size in range [4-150].  We study community detection from the following perspectives: 1) size of the detected communities, 2) percentage of users assigned to a desirable community, called the coverage of the communities, 3) extended modularity, 4) triangle participation ratio (TPR),  and 5) the interest of users in the same community. For user interest, we collect the top 10 hashtags tweeted by a user and manually inspect the hashtags to decide the user interest.  In addition, we propose a simple and intuitive community detection algorithm called Clique Augmentation Algorithm (CAA) which augment the cliques in the network into communities. We use growing threshold and overlapping threshold to control the size of the community and the amount of overlap among communities. 

We evaluated five widely used community detection algorithms on a Twitter topology of over 318,233 users we collected in 2013. 
Experimental results show that the well-accepted InfoMap~\cite{Infomap2008} outperforms Newman's Leading Eigenvector~\cite{EigenVector2006},  Fast Greedy~\cite{FastGreedy2004}, and Multilevel~\cite{MultilevelBlonde2008} algorithms in terms of community size distribution, community coverage, and user interest. For example, InfoMap assigns $62\%$ of users in the network into meaningful communities of size in the range of 4 to 150, while $93\%$ of the size of the communities generated by Eigenvector algorithm falls in the range of 1 to 3, which leads to less than $7\%$ of users are assigned to a meaningful community. We also observe that our proposed CAA algorithm produces communities of desired size and coverage. For example, $41\%$ of users are assigned to meaningful communities by CAA. In addition, CAA outperforms all other algorithms in Triangle Participation Ratio and demonstrates decent modularity. Finally, we show that the users in CAA communities show strongest similarity than communities obtained by all other algorithms. 

Our contributions in this paper are the following: 
\begin{enumerate}
\item We investigate an important but overlooked metric, the size of the community, to evaluate the quality of communities. To the best of our knowledge, this is the first paper that carry out empirically study on the size of the community and the modularity of community with different sizes. 
\item We discover that existing algorithms which optimize modularity can be significantly improved if considering community size during the optimization process
\item We investigate the community theme through hashtags
\item We demonstrate that a heuristic clique augmentation algorithm can produce high quality overlapping communities, which are needed by social media 
\end{enumerate}

The remainder of this paper is organized as follows. Section 2 describes related work in community detection algorithms and performance comparison. Section 3 introduces the proposed clique augmentation algorithm. Performance evaluation is given in Section 4 and Section 5 concludes this paper and outlines our future work.


\section{Related Work}
Much work has been conducted in the area of community detection along with ways to determine the quality of the identified communities. One comprehensive survey of recent advances \cite{Santo2010} discusses a wide range of existing algorithms including traditional methods, modularity based methods, spectral algorithms, dynamic algorithms, and more. Similarly, \cite{Yang2012} points out that it is important for the community detection algorithm to extract functional communities based on ground truth, where functional ground-truth community is described as a community in which an overall theme exists.

Another relevant paper \cite{Jierui2012} discusses overlapping community detection algorithms along with various quality measurements. Community detection often tries to optimize various metrics such as modularity as described by Girvan and Newman \cite{Newman2004} or conductance. The work in \cite{Jure2010} discusses many of the various objective functions currently in use and how they perform.

There exist many community detection algorithms in literature. We can categorize them into disjoint algorithms and overlapping algorithms, based on whether the identified communities have overlap or not. Infomap~\cite{Infomap2008} stands out as the most popular and widely used disjoint algorithm. The Infomap algorithm is based on random walks on networks combined with coding theory with the intent of understanding how information flows within a network. Multilevel~\cite{MultilevelBlonde2008} is a heuristic based algorithm based on modularity optimization. Multilevel first assigns every node to a separate community, then selects a node and checks the neighboring nodes attempting to group the neighboring node with the selected node into a community if the grouping results in an increase in modularity. Newman's Leading Eigenvector \cite{EigenVector2006} works by moving the maximization process to the eigenspectrum to maximize modularity by using a matrix known as the modularity matrix. Fast Greedy \cite{FastGreedy2004} is based upon modularity as well. It uses a greedy approach to optimize modularity. In the category of overlapping algorithm, Clique Percolation Method \cite{Palla2005} is the most prominent which merges two cliques into a community if they overlap more than a threshold.

\section{Clique Augmentation Algorithm}

In this section we propose a clique based algorithm to find communities and we call it Clique Augmentation Algorithm (CAA). CAA is built on the following two principles: 1) Users in a maximal clique belong to a stable community since a clique is densely connected internally;  2) A neighboring node that is highly connected to a clique should be part of the community since it keeps the triadic closure property among all nodes in the community. 


Given a social network topology, CAA algorithm discovers communities in the topology using the following steps: \\
\indent Step 1: Find all maximal cliques in the topology \\
\indent Step 2: Filter the overlapping cliques.  We sort the cliques based on their size then use an {\em overlapping threshold} to control the amount of overlap between two cliques.The overlapping threshold is defined as the percentage of overlapping nodes in the smaller clique. For example, given two cliques $c_{1}$ and $c_{2}$ where $c_{1}$ is of size 10, and $c_{2}$ is of size 5. Suppose the overlapping threshold is 0.7. If $c_{2}$ only has 2 nodes overlapping with $c_{1}$, we consider $c_{1}$ and $c_{2}$ as two independent cliques since $2 < 5*0.7$ which is 3.5. If $c_{2}$ had 4 overlapping nodes with $c_{1}$, we would discard $c_{2}$ since $4 > 3.5$.  \\
\indent Step 3: Grows each clique into a community by adding new nodes in one by one. {\em Growing threshold} is utilized for controlling the growth of each community. The growing threshold is defined as the ratio of the number of incoming edges from the new node to other nodes in the community over the size of a community. For example, if a community has size of 10, and the growing threshold is set to 0.7, then for a neighboring node to be added into the community, it must have at least 7 edges coming into the community. The algorithm checks the neighboring nodes for each node within the current community. This process is repeated for the updated community until no more nodes can be added. The growing threshold allows us to zoom into or out of the graph around the clique. \\

It is worth noting that CAA takes a different approach than Clique Percolation Method (CPM) where two adjacent cliques are merged into a community structure. In CAA, instead of merging neighboring cliques, we simply grow the community structure by adding individual node to the community sequentially. CAA has a few nice features: 1) it tends to be faster than CPM and manages to produce similar results. 2) CAA tries to capture the natural growth process of a community in the sense that if a user befriends with many users in a densely connected community, the user will be most likely grow as part of the community. 3) Experiment result shows CAA tends to produce decent sized communities and users in same community share strong interest.

\section{Experimental Results}

\subsection{Description of Data Set}
We carry out the comparative research on a Twitter Arizona user follower topology that we collected in summer 2013 which contains $318,233$ Arizona Twitter users with $3,545,258$ directed edges, and we call it AZTopology. An undirected graph is derived from AZTopology by removing all non-mutual edges and the isolated nodes. We call it AZTopologyMutual. AZTopologyMutual network contains $190,520$ nodes and $1,001,528$ undirected edges.

\subsection{Growing Threshold and Overlapping Threshold}
\begin{figure}[h]
  \includegraphics[width=\linewidth]{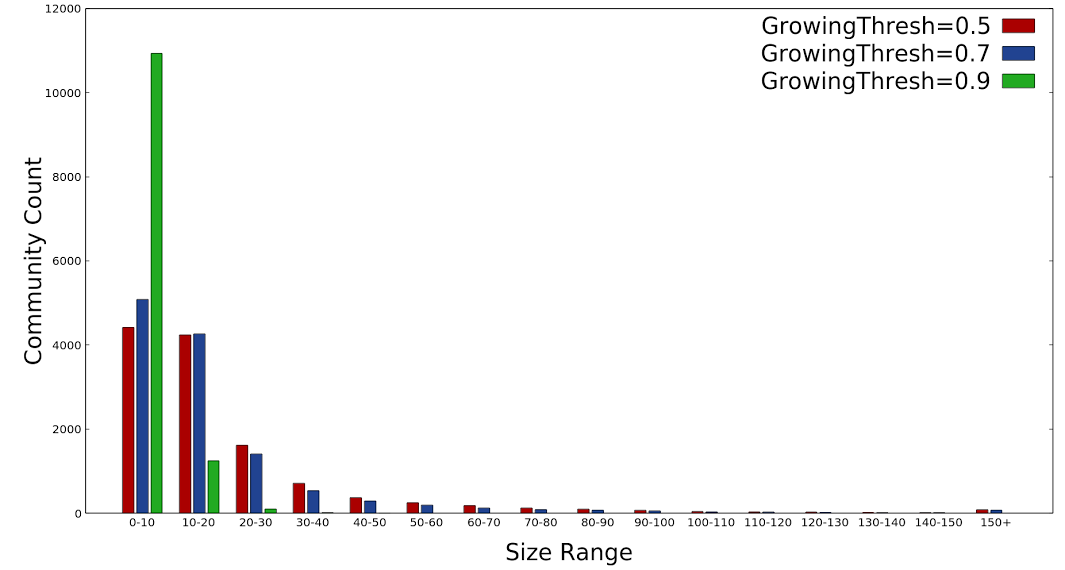}
  \caption{The Effect of Growing Threshold}
  \label{fig:growingThreshold}
\end{figure}

In this section, we measure the impact of growing threshold and overlapping threshold on community size and the number of communities in order to give suggestions on the parameter selection for CAA algorithm. 
First we look into the effect of growing threshold. We find all cliques of size 3 and larger in AZMutualTopology and set overlapping threshold to 0 to find all non-overlapping cliques. We set the growing threshold to 0.5, 0.7, and 0.9 indicating a neighboring node can only be added to the community if it is connected to at least 50\%, 70\%, or 90\% of nodes in the community. The result is plotted in Fig.~\ref{fig:growingThreshold}. x-axis is the community size range and y-axis is the number of communities whose size fall in the range. It can be seen that the cliques do not grow much with growing threshold $0.9$ since most communities are of size between 3 and 9. With growing threshold set to $0.7$, the number of communities with size in range [3-9] drops significantly, from around 11,000 to 5,000 and there are more communities in the range of [10-150]. Therefore, we recommend to set the growing threshold to $0.7$. Another interesting observation is that there is no significant difference in the distribution of community size for growing threshold $0.5$ and $0.7$. 

\begin{figure}[h]
	\centering
		\includegraphics[width=\linewidth]{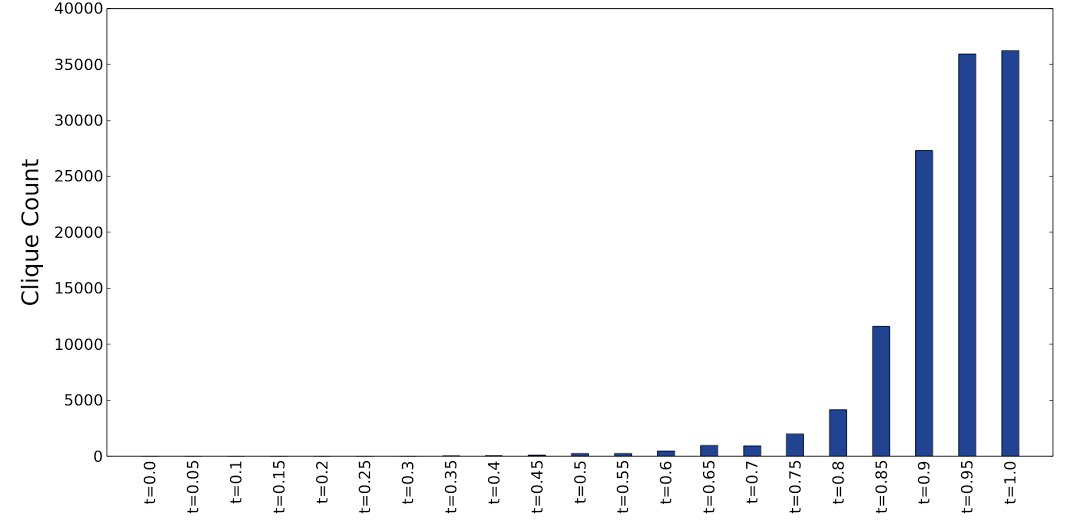}
	\caption{The Effect of Overlapping Threshold}
	\label{fig:overlappingThreshold}
\end{figure}

Next we investigate the effect of overlapping threshold. We choose all cliques of size over 15 and increase the overlapping threshold from 0 to 1. Intuitively, by increasing the overlapping threshold, less cliques are filtered, therefore the number of communities increases. As can been seen in Fig.~\ref{fig:overlappingThreshold}, where x-axis is the overlapping threshold value, y-axis is the number of cliques, the number of cliques increases significantly for overlapping threshold $\geq 0.8$. In general, we suggest to choose overlapping threshold less than 0.6 to avoid having heavily overlapping communities.  

\subsection{Comparison Of Community Detection Algorithms}

Previous survey papers~\cite{Harenberg2014} and \cite{Jure2010} have carried out comparative research on community detection algorithms and proposed different evaluation metrics. However, they ignored an important factor, that is, the size of the communities. Extremely large communities do not represent strong and stable communities. For this research we are interested in smaller communities where users are actually communicating and an overall theme amongst users exists. As described by Dunbar \cite{Dunbar2016}, the Dunbar number 150 applies to social media as well. This indicates that community size is an important factor and communities with more than 150 users are less desirable communities. 
On the other hand, communities of size 1, 2, and 3 are trivial. Therefore, we propose to study communities of size in [4-150] and we call such communities desirable communities. Users in such communities have stronger influence among each other and are less likely to leave the community. 

We propose to compare different community detection algorithms with the following criteria: community size, community coverage, extended modularity, triangle participation ratio, and the hash-tag similarity among users in the same community. We adopt the graph package for the implementation of Newman's Leading Eigenvector~\cite{EigenVector2006}, Infomap~\cite{Infomap2008}, Multilevel Algorithm~\cite{MultilevelBlonde2008}, Fast Greedy Optimization of Modularity~\cite{FastGreedy2004}, Label Propagation~\cite{LabelPropgatoin2007}, Edge Betweeness~\cite{EdgeBetweeness2002}, Clique Percolation Method (CPM)~\cite{Palla2005}, and implement our proposed CAA algorithm. The growing threshold of CAA is set to 0.7 and its overlapping threshold is set to 0. The communities grow from cliques of size $\ge 3$. Each algorithm was given AZTopologyMutual graph as the input and five hours to complete. Out of the eight algorithms, Newman's Leading Eigenvector, Infomap, Multilevel, Fast Greedy and CAA finished within five hours. So we only show the performance of these five algorithms. We run the algorithms on undirected graph to ensures fairness since not all algorithms support directed graph. Only Infomap, Label Propagation, Edge Betweenness, and CAA can run on directed graph. 

\subsubsection{Community Size} 
In this section, we present the number of communities and the size distribution of the communities. Table~\ref{tab:totalcomm} summarizes the number of communities and the size of the largest community revealed by each algorithm. Modularity maximization algorithms such as Multilevel, Eigenvector, and Fastgreedy all group users into large communities. For example, the larges community Eigenvector produces has size of 136,403, that is, over $70\%$ of all users are grouped into a large community. There is lack of strong connections among community users in such large community and we can hardly put this large community to practical use. 

\begin{table}[t]
\centering
\begin{tabular}{|c|c|c|} \hline
   Algorithm & Number of Communities & Largest Community Size\\ \hline
    CAA & 12,312 & 680\\ \hline
    Infomap & 18,537 & 13,126 \\ \hline
    Multilevel & 7,409 & 34,955 \\ \hline
    Eigenvector & 5,834 & 136,403\\ \hline
    Fastgreedy & 9,350 & 51,781\\ \hline
\end{tabular}
\caption{Total number of detected communities and the size of largest community}
\label{tab:totalcomm}
\end{table}

\begin{figure}[h]
		\centering
	\subfigure[Community Size Distribution from 1 to 50] {
		\centering
		\includegraphics[width=3.5in]{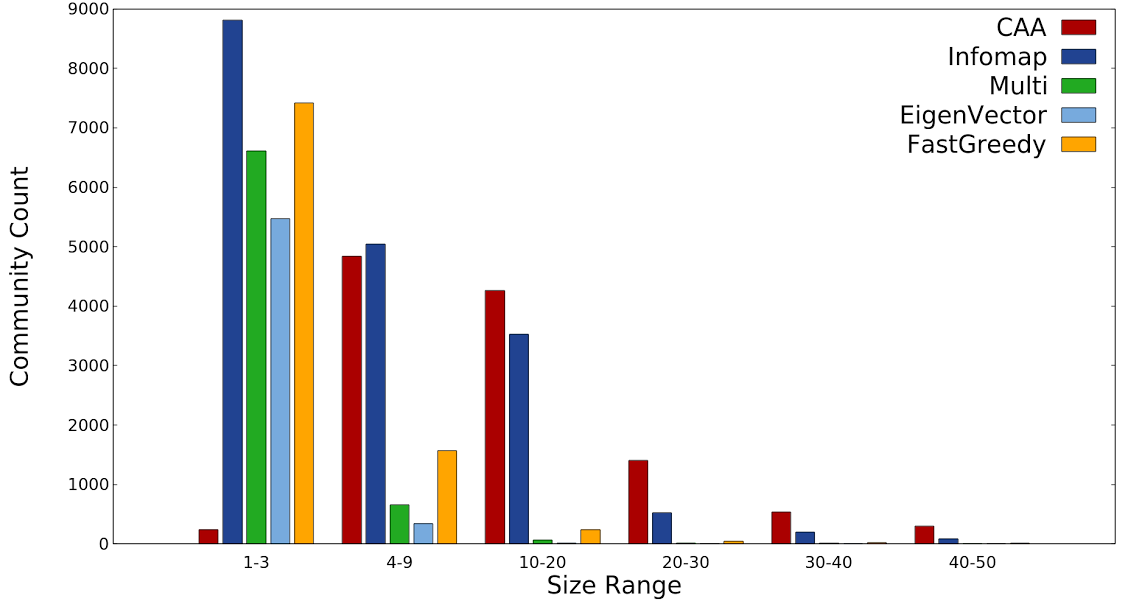}}	
	\subfigure[Community Size Distribution from 50 to 150+] {
		\centering
		\includegraphics[width=\linewidth]{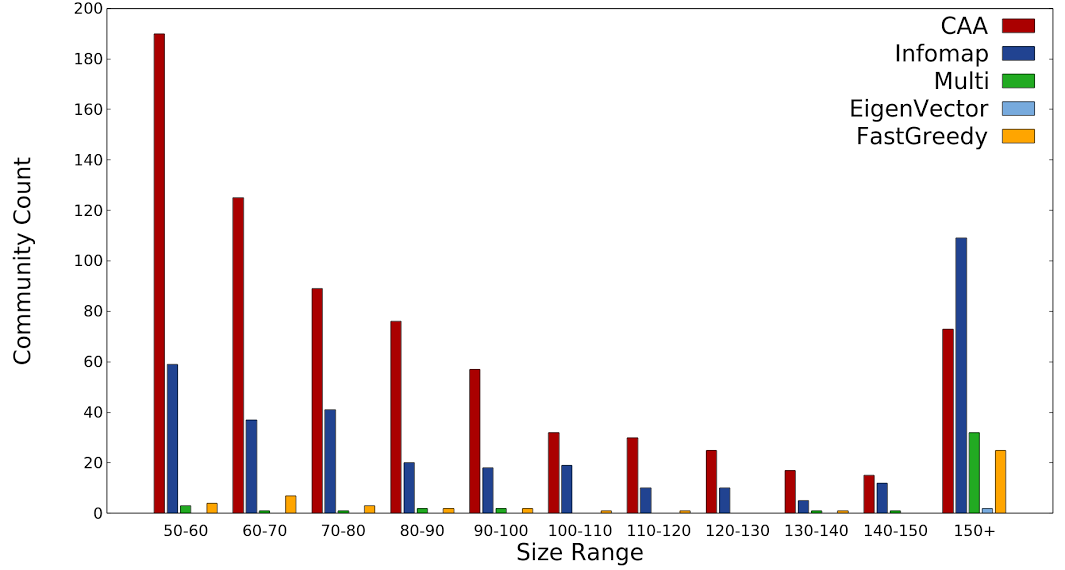}}
	\subfigure[Community Size Percentage Distribution] {
		\centering
		\includegraphics[width=\linewidth]{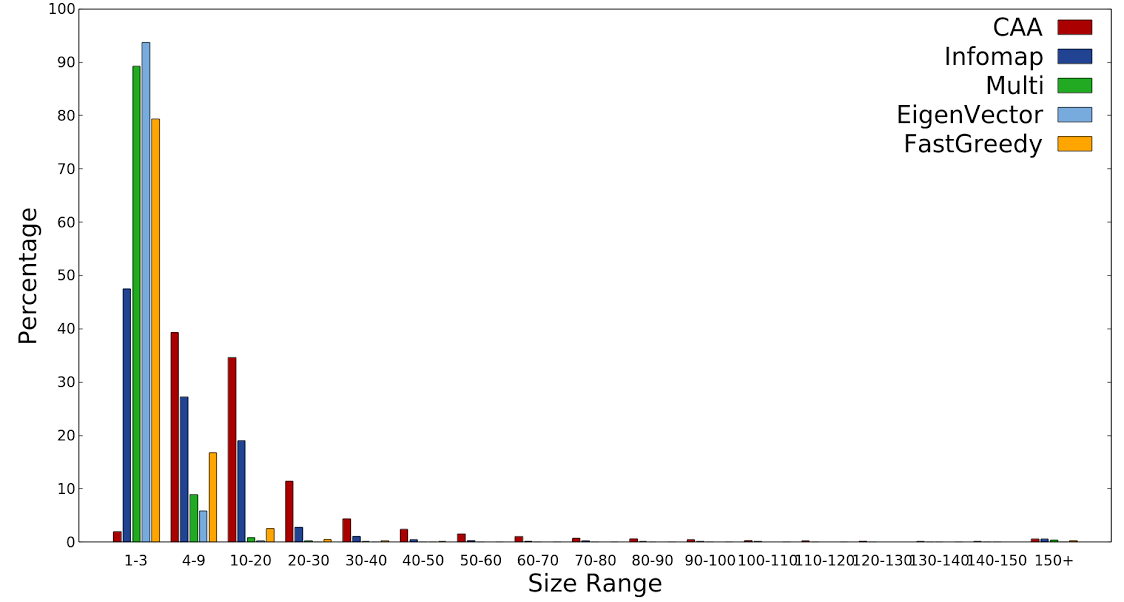}}
	\caption{The Size Distribution of Communities}
	\label{fig:distribution}
\end{figure}

We plot the community size distribution of communities obtained by CAA, Infomap, Multilevel, EigenVector, and FastGreedy in Fig.~\ref{fig:distribution}. Due to the significant difference in the scale of the number of communities with difference sizes, we split the result into two subgraphs. Fig.~\ref{fig:distribution}(a) illustrates the number of detected communities whose size fall in the range of 1 to 50 and Fig.~\ref{fig:distribution}(b) illustrates the number of detected communities with size in the range above 50. As discussed, desirable communities have size in the range of 4 to 150. It is clear that both CAA and InfoMap have decent number of desirable communities. Furthermore, CAA produces the largest number of desirable communities. Multilevel produces 11 communities in the range of 50 to 150, FastGreedy obtains 21 communities in this range, and EigenVector produces 0 communities in this range. Fig.~\ref{fig:distribution}(c) illustrates the percentage of communities with size in a certain range. It is conclusive that CAA and InfoMap can be used to obtain desirable communities in social media. CAA outperforms InfoMap in terms of community size since over $95\%$ of all communities produced by CAA are desirable communities and less than $55\%$ of all communities produced by InfoMap are desirable communities. For all three other algorithms, majority of their communities have size in the range of [1-3] . Moreover, the largest community has extreme size, therefore, there are less than $5\%$ desirable communities. 

\subsubsection{Community Coverage} 
\begin{figure}[h]
  \includegraphics[width=\linewidth]{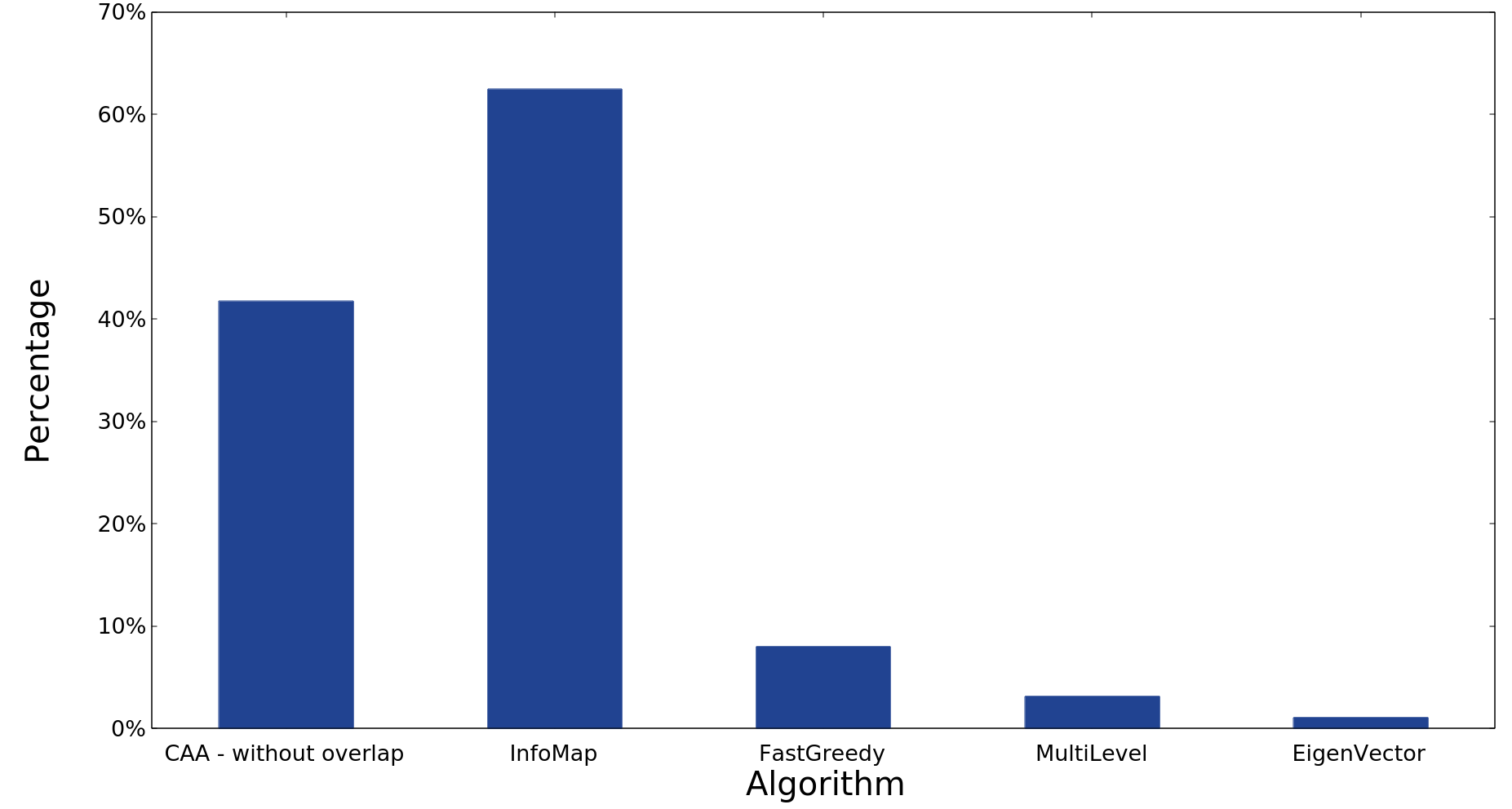}
  \caption{Desirable Community Coverage [4, 150]}
  \label{fig:coverage}
\end{figure}

It is desirable that a community detection algorithm can produce communities that cover the whole network, that is, everybody in the network is assigned to a community. Meanwhile, the size of the community matters. Too small community or too big community does not provide much value if the goal is to find strong communities. So we calculate the community coverage of each detection algorithm by only considering the users assigned to communities of size between 4 and 150. Even InfoMap, Multilevel, Eigenvector, and FastGreedy all assign every single node in the graph to a community, they don't provide $100\%$ coverage in our calculation. Fig.~\ref{fig:coverage} shows the performance of each algorithm with regard to community coverage while taking the community size into account.  

It is clear that Infomap performs the best in terms of community coverage since it assigns $62\%$ of all users in the network into a desirable community. CAA also performs remarkably well when compared to other methods. $41\%$ of users are assigned to desirable communities by CAA. It is worth noting that in this experiment, CAA starts from non-overlapping cliques by setting the overlapping threshold to 0. In fact, CAA can achieve higher coverage if we allow overlapping cliques. FastGreedy, MultiLevel, and EigenVector failed to assign majority of users in the network to meaningful communities. For example, 93\% of the size of the communities generated by Eigenvector algorithm falls in the range of 1 to 3, which leads to less than 7\% of users assigned to a desirable community. 

\subsubsection{Extended Modularity} 
Extended modularity has been defined as a popular metric to measure the goodness of overlapping communities in~\cite{Huawei2009}. We give the definition of this metric in Equation~\eqref{eq:extendedModularity}, where $O_{v}$ is the number of communities the vertex $v$ belongs to, similarly $O_{w}$ is the number of communities vertex $w$ belongs to. $A$ is the adjacency matrix, that is, $A_{vw} = 1$ means there exists an edge between vertex $v$ and vertex $w$, otherwise it is 0. $\frac{k_v k_w}{2m}$ describes the expected number of edges between vertex $v$ and vertex $w$. $k_v$ and $k_w$ are the node degree of $v$ and $w$ in the whole topology respectively. $m$ is the total number of edges in the whole topology. The range of extended modularity is $[-1,1]$ and the higher the value, the better the community in terms of modularity. The intuitive behind extended modularity is that communities should have more internal connectivity than random graph of the same degree sequence. Note that it is called extended modularity since it is a variation to modularity, which is the most widely used metric to measure the goodness of disjointed community. To be more specific, if there is no overlap between the communities, $\frac{1}{O_v O_w}$ is 1 and Equation \eqref{eq:extendedModularity} becomes the traditional modularity as defined in~\cite{Clauset2004}. In general, overlapping communities have lower modularity than disjoint communities since overlapping communities have many connections outside the community.
\begin{displaymath}
  \displaystyle EQ = \frac{1}{2m} \sum_{i} \sum_{v \in C_i, w \in C_i }  \frac{1}{O_v O_w}\Bigl[ A_{vw} - \frac{k_v k_w}{2m} \Bigl] 
  \tag{1}
  \label{eq:extendedModularity}
\end{displaymath}

\begin{figure}[h]
  \includegraphics[width=\linewidth]{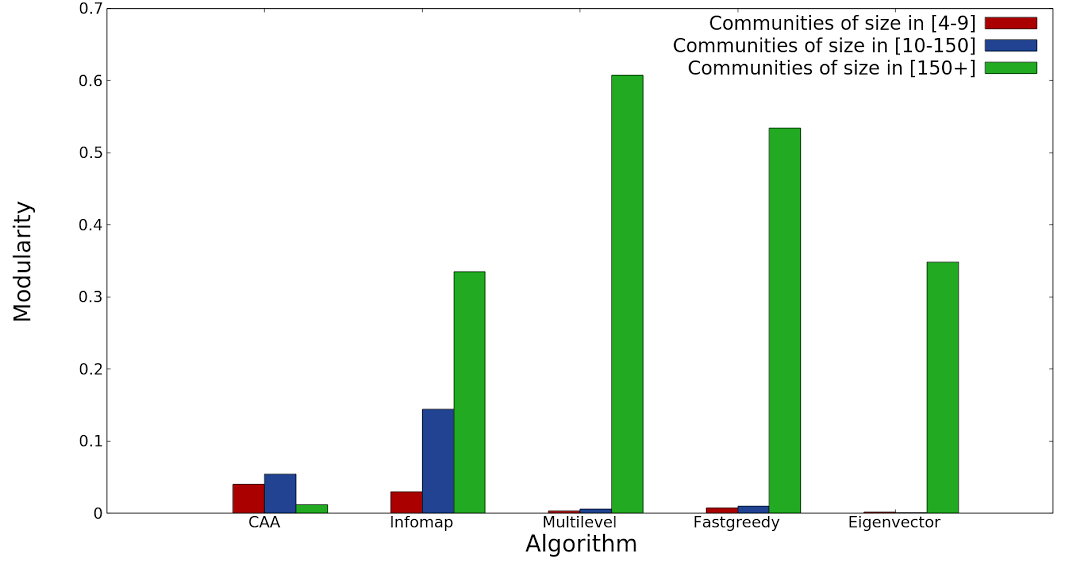}
  \caption{Extended Modularity}
  \label{fig:extendedModularity}
\end{figure}

In order to have a closer look at how communities of different sizes contribute to the modularity, we divide the sum over all communities in Equation \eqref{eq:extendedModularity} into three parts: the sum over all communities of size in range [4-10), the sum over communities of size in range [10-150),  and the sum over communities of size in range  [150+] respectively. Figure~\ref{fig:extendedModularity} presents how communities of different sizes contribute to the extended modularity value. An interesting discovery is that even modularity-based algorithm Multilevel, FastGreedy, and EigenVector achieve high modularity in general, the communities of size in the range of 4 to 150, which is desirable communities, contribute surprisingly low to the high modularity. These algorithms show strong modularity because the communities with large size have high modularity. The result clearly shows that for all four algorithms except CAA, the contribution of communities with large size to the calculation of extended modularity are significantly higher than that of communities with desirable size. This can be partially attributed to the resolution limit of modularity~\cite{Fortunato2007}. This discovery has a significant impact since it suggests that we can design communities detection algorithms to discover strong communities if we take both modularity and size into consideration.

\subsubsection{Triangle Participation Ratio}

\begin{figure}[h]
  \includegraphics[width=\linewidth]{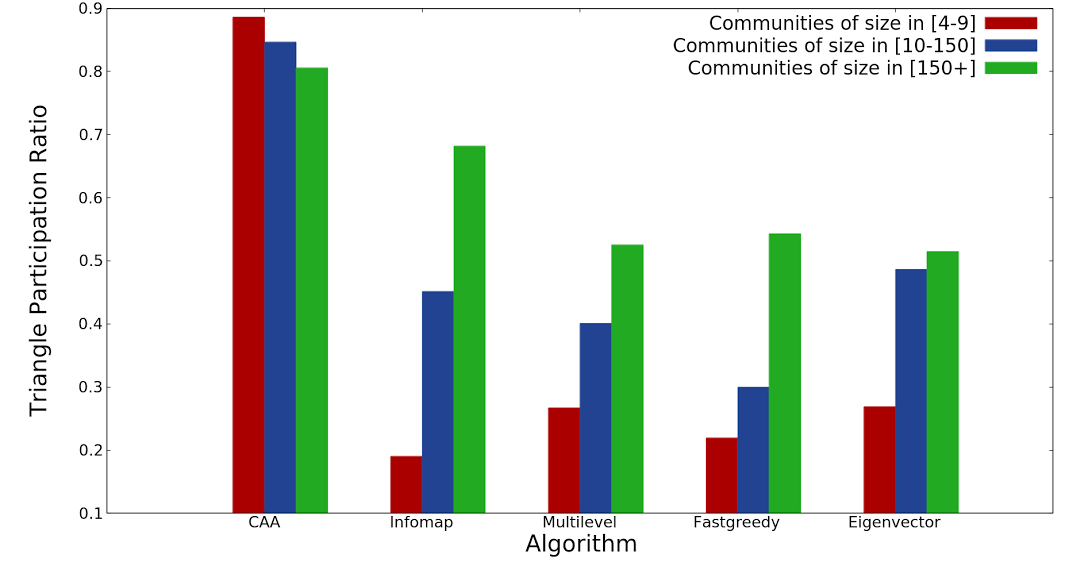}
  \caption{Triangle Participation Ratio}
  \label{fig:tpr}
\end{figure}

Triangle Participation Ratio (TPR) was proposed in~\cite{Yang2012} as a metric for community evaluation, where it is defined as the number of nodes in a community that form a triad, divided by the total number of nodes in the community.  ~\cite{Yang2012} found that the communities discovered by algorithms to optimize community modularity don't align with the ground truth communities, that is, the known membership in real life, while TPR is a good metric when looking for ground-truth communities. In this section, we plot the TPR of communities of different size groups: [4-9], [10-150], [150+], generated by all five algorithms in Fig.~\ref{fig:tpr}. As indicated by the figure, CAA achieves significantly higher TPR than other four algorithms. This is expected since CAA starts from a clique with TPR as 1 and grows the clique with new nodes that are highly connected to the growing community. 

\subsubsection{Hash-Tag Similarity}
People form communities for collaboration and communication. It is natural that users in the same community share some interest. In this section, we compare the performance of community detection algorithms in terms of the similarity of users in the same community. To be more specific, we randomly pick 50 communities with size in the range of [10-150] and collect the tweets for all users in the community, then for each user, find their top 10 hashtags. We consider a community as a ``good community'' if there appears to be a common theme amongst all members in the community or a common hash-tag usage by all members. Table~\ref{tab:goodcomm} shows the number of good communities based on user hashtag similarity. For EigenVector there are less than 50 communities to choose from within the reasonable range therefore we analyze all 20 communities in this range. In general, we find most of the algorithms are poor at detecting communities in which a theme exists.

\begin{table}[h]
\centering
\begin{tabular}{ | l | l | l |} \hline
    Algorithm & Good Communities \\ \hline
    CAA & 10  \\ \hline
    Infomap & 7  \\ \hline
    Multilevel & 6 \\ \hline
    Eigenvector & 3  \\ \hline
    Fastgreedy & 7 \\ \hline
\end{tabular}
\caption{User Hashtag Similarity Comparison}
\label{tab:goodcomm}
\end{table}

\begin{table}[h]
\begin{tabular}{ | c | } \hline
\parbox[t]{8.27cm} { \centering Top Hash Tags In A Good Community} \\
\end{tabular} \\
\begin{tabular}{ | c | c | } \hline
User 1 & \parbox[t]{6.5cm}{ \#PJNET 77, \#USArmy 74, \#Pray 51, \#UnbornLivesMatter 45, \#tcot 31, \#Catholic 26, \#Jesus 25, \#Trump2016 25, \#ChooseLife 22, \#Brexit 19} \\ \hline
User 2 & \parbox[t]{6.5cm}{\#GOPDebate 166, \#Trump 82, \#tcot 73, \#Hillary2016 37, \#gop 35, \#ccot 34, \#PJNET 33, \#CruzCrew 30, \#Obamigration 30, \#HillaryEmail 29
} \\ \hline
User 3 & \parbox[t]{6.5cm}{ \#Trump2016 157, \#Trump 76, \#tcot 50, \#TrumpTrain 50, \#MAGA 44, \#MakeAmericaGreatAgain 37, \#PJNET 33, \#RetireMcCain 32, \#NeverHillary 25, \#CruzSexScandal 24 } \\ \hline
Community & \parbox[t]{6.5cm}{\#tcot 6070, \#ocra 3018, \#tlot 2062, \#football 1359, \#sportinggoods 1307, \#sports 1295, \#PJNET  887,  \#Thanks 855, \#tgdn 774, \#TGDN 759, \#Chesterton 743, \#pjnet 681, \#Benghazi 678, \#GOPDebate 629, \#Lauds 616, \#AZ 475, \#CruzCrew 400, \#ccot 366, \#Trump2016 341, \#Trump 329}\\ \hline
\end{tabular}
\caption{An example of user hashtags of a good community}
\label{tab:goodexample}
\end{table}

Table~\ref{tab:goodexample} shows the top 10 hashtags used by three users and top 20 hashtags used by the whole community in a politics community. All hashtags were collected in July, 2016. The community is produced by CAA and has 48 users in it. We also list the number of times each hashtag was used next to the hashtag. By analyzing this community hashtags, we can tell it is a conservative political community. We have also discovered communities with other themes such as restaurant, wine, and Sedona travel. 

\begin{table}[h]
\begin{tabular}{ | c | c | } \hline
    Community Size & Top 10 Hash Tags of The Community \\ \hline
    167 & \parbox[t]{6cm}{ \#tbt, \#beardown, \#blessed, \#tuscon, \#wtf, \#truth, \#lancernation \#wcw, \#yolo, \#fml} \\ \hline
    276 & \parbox[t]{6cm}{ \#tbt, \#wcw, \#mcm, \#beardown, \#truth, \#sorrynotsorry, \#thestruggle \#subtweet, \#smh, \#ifwedate} \\ \hline
    281 & \parbox[t]{6cm}{ \#wcw, \#tbt, \#ifwedate, \#mcm, \#oomf, \#sorrynotsorry, \#blessed, \#rt, \#retweet, \#truth } \\ \hline
\end{tabular}
\caption{Top 10 Hashtags of Large Sized Communities}
\label{tab:badcomm}
\end{table}

Additionally, we inspect large communities of size between 150-300 to check whether these communities make sense. We did this for the InfoMap method because this is one of the most popular community detection algorithms. Table~\ref{tab:badcomm} shows a sampling of three randomly selected communities. Here the hashtags are sorted so that the top hashtag appears first.


By interpreting the meaning of these hashtags, where \#wcw stands for Women-Crush Wednesday, \#mcm stands for Man-Crush Monday, \#smh stands for Shake My Head, and \#oomf stands for One Of My Followers, we find that each individual communities in general does not make much sense. This is because the hashtags we found are everyday used terms on Twitter. As such, the communities lack an overall theme.  This is consistent to what we expected to see at larger community sizes.

\section{Conclusion}

In this paper, we study the problem of evaluating community detection algorithm by introducing three new measurement, the community size, community coverage and user interest. We propose a simple clique based algorithm CAA as baseline and compare the performance of four popular algorithms.  CAA discovers overlapping communities, therefore is a good fit for social media.  Our findings indicate that both InfoMap and CAA are able to discover desirable communities which consist of users sharing similar interests, while many existing algorithms generate too small or too big communities. We plan to automate the user interest labeling by adopting topic modeling in our future work. We also plan to design new algorithms to  maximize modularity while consider the size of the community. 

\bibliographystyle{IEEEtran}


\begin{thebibliography}{10}
\providecommand{\url}[1]{#1}
\csname url@rmstyle\endcsname
\providecommand{\newblock}{\relax}
\providecommand{\bibinfo}[2]{#2}
\providecommand\BIBentrySTDinterwordspacing{\spaceskip=0pt\relax}
\providecommand\BIBentryALTinterwordstretchfactor{4}
\providecommand\BIBentryALTinterwordspacing{\spaceskip=\fontdimen2\font plus
\BIBentryALTinterwordstretchfactor\fontdimen3\font minus
  \fontdimen4\font\relax}
\providecommand\BIBforeignlanguage[2]{{%
\expandafter\ifx\csname l@#1\endcsname\relax
\typeout{** WARNING: IEEEtran.bst: No hyphenation pattern has been}%
\typeout{** loaded for the language `#1'. Using the pattern for}%
\typeout{** the default language instead.}%
\else
\language=\csname l@#1\endcsname
\fi
#2}}

\bibitem{Dunbar2016}
     Dunbar, R. I. M. ``Do Online Social Media Cut through the Constraints That Limit the Size of Offline Social Networks? `` in \emph {R. Soc. Open Sci. Royal Society Open Science 3.1 (2016): 150292.}
\bibitem{Ding2016}
	Ding, Z. et al. ``Overlapping Community Detection based on Network Decomposition.`` in \emph {Sci. Rep. 6, 24115; doi: 10.1038/srep24115 (2016).}
\bibitem{Harenberg2014}
     Harenberg, Steve, Gonzalo Bello, L. Gjeltema, Stephen Ranshous, Jitendra Harlalka, Ramona Seay, Kanchana Padmanabhan, and Nagiza Samatova.  ``Community Detection in Large-scale Networks: A Survey and Empirical Evaluation. `` in \emph {Wiley Interdisciplinary Reviews: Computational Statistics WIREs Comput Stat 6.6 (2014): 426-39.}
\bibitem{Yang2012}
     Yang, Jaewon, and Jure Leskovec. ``Defining and Evaluating Network Communities Based on Ground-Truth.`` in \emph {2012 IEEE 12th International Conference on Data Mining (2012). }
\bibitem{Jierui2012}
	Xie, Jierui, Stephen Kelley, and Boleslaw K. Szymanski. ``Overlapping Community Detection in Networks.`` in \emph {CSUR ACM Comput. Surv. ACM Computing Surveys 45.4 (2013): 1-35.}
\bibitem{Jure2010}
	Leskovec, Jure, Kevin J. Lang, and Michael Mahoney. ``Empirical Comparison of Algorithms for Network Community Detection`` in \emph {Proceedings of the 19th International Conference on World Wide Web - WWW '10 (2010).}
\bibitem{Santo2010}
	Fortunato, Santo. ``Community Detection in Graphs.`` in \emph {Physics Reports 486.3-5 (2010): 75-174.}
\bibitem{Lazar2010}
	Lazar, A., D. Abel, and T. Vicsek. ``Modularity Measure of Networks with Overlapping Communities.`` in \emph {EPL (Europhysics Letters) Europhys. Lett. 90.1 (2010): 18001.}
\bibitem{Huawei2009}
	Shen, Huawei, Xueqi Cheng, Kai Cai, and Mao-Bin Hu. ``Detect Overlapping and Hierarchical Community Structure in Networks.`` in \emph {Physica A: Statistical Mechanics and Its Applications 388.8 (2009): 1706-712.}
\bibitem{Infomap2008}
     Rosvall, M., and C. T. Bergstrom. ``Maps of random walks on complex networks reveal community structure,'' in \emph{Proceedings of the National Academy of Sciences 105.4.}, 2008.
\bibitem{MultilevelBlonde2008}
     Blondel, Vincent D., Jean-Loup Guillaume, Renaud Lambiotte, and Etienne Lefebvre. ``Fast Unfolding of Communities in Large Networks.'' in \emph{J. Stat. Mech. Journal of Statistical Mechanics: Theory and Experiment.}, 2008.
\bibitem{Fortunato2007}
	Fortunato, S., and M. Barthelemy. ``Resolution Limit in Community Detection.'' in \emph{Proceedings of the National Academy of Sciences 104.1 (2006): 36-41.}
\bibitem{LabelPropgatoin2007}
	Raghavan, Usha Nandini, Réka Albert, and Soundar Kumara. ``Near Linear Time Algorithm to Detect Community Structures in Large-scale Networks.`` in \emph {Physical Review E Phys. Rev. E 76.3 (2007).}
\bibitem{Java2007}
	Java, Akshay, Xiaodan Song, Tim Finin, and Belle Tseng. ``Why We Twitter.`` in \emph {Proceedings of the 9th WebKDD and 1st SNA-KDD 2007 Workshop on Web Mining and Social Network Analysis - WebKDD/SNA-KDD '07 (2007).}
\bibitem{EigenVector2006}
	Newman, M. E. J. ``Finding Community Structure in Networks Using the Eigenvectors of Matrices.`` in \emph {Physical Review E Phys. Rev. E 74.3 (2006).}
\bibitem{Palla2005}
	Palla, Gergely, Imre Derényi, Illés Farkas, and Tamás Vicsek. ``Uncovering the Overlapping Community Structure of Complex Networks in Nature and Society.'' in \emph {Nature 435.7043 (2005): 814-18.}
\bibitem{FastGreedy2004}
	Newman, M. E. J. ``Fast Algorithm for Detecting Community Structure in Networks.`` in \emph {Physical Review E Phys. Rev. E 69.6 (2004).}
\bibitem{Newman2004}
	Newman, M.E.J., Girvan M. ``Finding and evaluating community structure in networks.`` in \emph {Phys. Rev. E. 69, 026113 (2004).}
\bibitem{Clauset2004} 
	A. Clauset, M. E. J. Newman, and C. Moore ``Finding community structure in very large networks.`` in \emph {Phys. Rev. E. 70, 066111 (2004).}	
\bibitem{EdgeBetweeness2002}
	Girvan, M., and M. E. J. Newman. ``Community Structure in Social and Biological Networks.`` in \emph {Proceedings of the National Academy of Sciences 99.12 (2002): 7821-826.}

\end{thebibliography}
\end{document}